\documentclass{article}
\usepackage{graphicx} 
\usepackage{color, soul}
\usepackage{cite}
\usepackage{subcaption}
\usepackage{ulem}
\usepackage{authblk}
\usepackage{changes}
\usepackage{verbatim}
\usepackage[a4paper, margin=2.7cm]{geometry}
\usepackage{hyperref}

\title{
{Polytropic Stars in $f(Q) = Q +\xi Q^2$ covariant formulation}}

\author[1]{J.C.N. de Araujo\thanks{jcarlos.dearaujo@inpe.br}}
\author[2]{H.G.M. Fortes\thanks{hemily.gomes@gmail.com}}
\affil[1]{Divisão de Astrof\'{i}sica, Instituto Nacional de Pesquisas Espaciais, Avenida dos Astronautas 1758, S\~{a}o Jos\'{e} dos Campos, SP 12227-010, Brazil}
\affil[2]{Departamento de Física de Volta Redonda, Universidade Federal Fluminense, Campus Aterrado, Rua Desembargador Ellis Hermydio Figueira 783, Volta Redonda, RJ, 27213-145, Brazil.}

\date{}

\begin{document}

\maketitle

\begin{abstract}
General Relativity (GR) is not the only way gravity can be geometrized. Instead of curvature, the Teleparallel Theory attributes gravity to the torsion $T$, which is related to the antisymmetric part of the connection. On the other hand, the Symmetric Teleparallel Theory no longer preserves metricity, describing gravity through the non-metricity tensor $Q_{\alpha\mu\nu}\equiv \nabla_\alpha g_{\mu\nu}.$ These descriptions give form to what is known as geometrical trinity of gravity. Recently, the extensions of GR have been intensively investigated in order to solve the theoretical impasses which have arisen. In this sense, it is also useful to investigate the extensions of alternative descriptions of gravity, which leads us to the so-called $f(T)$ and $f(Q)$ gravities. In this paper, we consider a family of $f(Q)$ models and obtain their corresponding Tolman-Oppenheimer-{Volkoff} equations applied to {polytropic} stars. It is worth mentioning that the $f(Q)$ covariant formulation is adopted and we present a new equation that substantially simplifies the numerical calculations.
 Using numerical integration, it is possible to solve a system of differential equations to calculate, among other things, the maximum mass and mass-radius relation allowed. In addition, we explicitly show the non-metricity behavior inside and outside the star.

\end{abstract}

\section{Introduction}
\label{int}

There are different descriptions of gravity in the literature, some entirely equivalent \cite{Capozziello2}. General relativity (GR) is the most successful theory of gravity ever proposed based on a geometric description through Riemannian geometry. Its triumph is due to accurate predictions for experimental and observational tests \cite{Will2}. However, despite its great success, the theory has been found insufficient to fully explain certain cosmological observations, such as the accelerated expansion of the universe \cite{Riess} and the presence of an unknown component called dark energy \cite{Peebles}.

However, in order to circumvent this impasse, it is natural to pursue new ways to describe gravity without disregarding the achievements of GR. In this sense, extensions of GR have been extensively studied in recent decades. The $f(R)$ models \cite{fR}, for instance, is a formulation in which the Einstein-Hilbert action is replaced for a more general one with an arbitrary function of the scalar curvature $R$. There are several works on $f(R)$ and its implications. However, the field equations of these models are in general fourth-order differential equations, which makes their analysis difficult. On the other hand, an alternative approach can also be considered, the so-called Teleparallel Theory of Gravity \cite{Aldro,review2,TT} is known to be equivalent to GR, with the basic difference of having in its formulation the torsion scalar $T$ instead of the scalar curvature $R$. In the same way that the General Relativity can be extended through the $f(R)$ models, the natural extension for the Teleparallel Theory of Gravity is the $f(T)$ models \cite{Ferraro,Linder} which have the advantage of leading to second-order field equations, therefore, simpler to analyze compared to the fourth-order equations of $f(R)$. In addition, $f(T)$ has presented interesting cosmological and astrophysical solutions, providing alternative interpretations of the acceleration phases of the universe \cite{Ferraro,Linder,Myr,Karami} and applications to compact stars \cite{Ganiou,Kpa,Pace,Ilijic,Bohmer,Pace2,FA1,FA2,FA3,FA4,FA5}. Note that although GR and Teleparallel Theory are completely equivalent, this is not necessarily true for their extended versions, $f(R)$ and $f(T)$, respectively. For this reason, it is important to not disregard the possibility of relevant results from those alternative models.

On the one hand, while TEGR and its extensions, such as the $f(T)$ models, differ from GR due to the inclusion of the antisymmetric part of the connection -- namely, torsion $T$ being nonzero -- on the other hand, both preserve metricity, i.e., $\nabla_\alpha g_{\mu\nu}=0$. In this sense, there should be a third possible description of gravity, the so-called Symmetric Teleparallel Equivalent of GR (STEGR), where there is no longer metricity and a new object is defined, namely, the non-metricity tensor $Q_{\alpha\mu\nu}\equiv \nabla_\alpha g_{\mu\nu}$. Therefore, in STEGR the action is given in terms of non-metricity and it completely reproduces the dynamics of GR. For a complete review of these three different geometrical descriptions of gravity, see \cite{Jimenez2}, where the authors refer to them as the {\it geometrical trinity of gravity}.

Similarly to how $f(R)$ and $f(T)$ models extend General Relativity and Teleparallel Gravity, the Symmetric Teleparallel Equivalent of GR can be extended by considering more general functions of nonmetricity in the action giving rise to the so-called $f(Q)$ gravity, which has been considered in many works recently \cite{Pradhan,Mandal,Bhardwaj,Myrzakulov,Adak,Zhang,Ambrosio,Blixt,Ferreira,Calza,Maurya,Lohakare,Shekh,Capozziello,Heisenberg,Errehymy,Capozziello2,Lin,Gakis,Zhao,Sokoliuk,Wang,Jimenez,Bhar,De,Pradhan2,Gadbail2}.

{In $f(Q)$ gravity, the action is constructed as a general function of the nonmetricity scalar $Q$, which characterizes the deviation from the Riemannian geometry in the symmetric teleparallel framework. This approach has gained significant attention because of its ability to describe gravitational interactions without torsion or curvature, offering a novel perspective on modified gravity theories. On the theoretical front, progress has been made in understanding the covariant formulation of f(Q) gravity \cite {Zhao} and its Hamiltonian structure, revealing challenges in the canonical quantization of the theory \cite{Ambrosio}. Recent studies have explored various aspects of $f(Q)$ gravity, including its cosmological implications, exact solutions, and astrophysical applications. For instance, investigations on cosmological models in $f(Q)$ gravity have been extensively studied, including the behavior of the cosmological perturbations and some accelerating solutions with relevance for inflation and dark energy \cite{Jimenez}. A comparison between these new cosmologies and the $\Lambda CDM$ set-up can be found in \cite{Ayuso}.
In addition, recent studies have begun to test these models against observational data, such as standard sirens, to constrain their parameters and validate their predictions \cite{Ferreira}. Additionally, exact solutions for static and spherically symmetric configurations have been derived, providing insights into the behavior of compact objects such as neutron stars and quark stars within this framework \cite{Calza,Lin,Sokoliuk,Wang}. The theory has also been extended to include boundary terms $f(Q,B))$ and couplings with electromagnetic fields \cite{Capozziello,Shekh}. Finally, the relationship and connections between f(Q) gravity and other modified gravity theories, such as f(R) and f(T), have been explored, emphasizing the unique features of nonmetricity-based theories and their potential to address open questions in cosmology and astrophysics \cite{Capozziello2}. 
}

{In this paper, we explore the implications of the simplest functional form for f(Q) gravity in the context of compact star models. Our analysis is structured as follows. In Section \ref{be}, we present the basic equations for extended f(Q) theories, with a particular focus on the modifications to the Tolman-Oppenheimer-Volkoff (TOV) equations in a spherically symmetric spacetime, providing the theoretical framework for our study. In Section \ref{sectionMS}, we specify the functional form of f(Q) and analyze both interior and exterior solutions for a spherically symmetric matter distribution. This section includes a discussion of the boundary conditions and the matching of the interior and exterior metrics, ensuring a consistent description of the stellar structure. In Section \ref{Ne}, we provide numerical examples using polytropic equations of state, illustrating the effects of f(Q) gravity on the mass-radius relations and other properties of compact stars. These numerical results are compared with those from General Relativity, highlighting the distinctive features of f(Q) gravity. Finally, in Section \ref{Fr}, we summarize our findings, discuss their implications for the understanding of compact objects in modified gravity, and outline potential directions for future research. Through this work, our aim is to explore f(Q) gravity in the context of compact stars, offering new insights into the behavior of matter under extreme conditions in alternative gravitational theories.}

\section{The basic equations of f(Q) gravity for spherically symmetric metric}
\label{be}

This section introduces the main equations of Symmetric Teleparallel General Relativity (STGR) and its extended theory, $f(Q)$. In STGR, gravity is described by the nonmetricity tensor, as both curvature and torsion are assumed to vanish.

{The nonmetricity scalar is defined as}
\begin{equation}
    Q = - P^{\alpha\beta\gamma} Q_{\alpha\beta\gamma},
\end{equation}
where $Q_{\alpha\beta\gamma}$ (the nonmetricity tensor) and $P^{\alpha\beta\gamma}$ (the nonmetricity conjugate, also known as the superpotential) are given by
\begin{equation}
    Q_{\alpha\beta\gamma} = \nabla_\alpha g_{\beta\gamma}
\end{equation}
and
\begin{equation}
    {P^{\alpha}}_{\beta\gamma} = -\frac{1}{4} {Q^{\alpha}}_{\beta\gamma} + \frac{1}{2}{Q_{(\beta\gamma)}}^\alpha + \frac{1}{4} (Q^\alpha - \tilde{Q}^\alpha)g_{\beta\gamma}-\frac{1}{2}\delta^\alpha_{(\beta}Q_{\gamma )},
\end{equation}
where $Q_\alpha \equiv {{Q_\alpha}^\mu}_\mu $ and $\tilde{Q}_{\alpha} \equiv {Q^{\mu}}_{\alpha\mu} $.

{The action for the extended theory $f(Q)$ reads}
\begin{equation}
S = \int   \ \left( \frac{f(Q)}{16\pi} + \mathcal{L}_m \right) \ \sqrt{-g}\ d^4 x \ .
\end{equation}
It is worth mentioning that in this paper we adopt the covariant formulation of $f(Q)$. For a detailed discussion, we refer the reader to Ref. \cite{Zhao}. {This distinction is crucial because, when using the coincident gauge where the affine connection components vanish, an issue arises inadvertently leading to the linear model in $Q$, thus recovering only General Relativity (GR). However, in \cite{Zhao}, a solution to this problem is provided by introducing a covariant formulation for $f(Q)$, which allows a non-vanishing connection and supports non-linear extensions in Q. In the same work, the covariant formulation is applied to spherically symmetric spacetimes, yielding general equations of motion f(Q). In the literature, many authors have used the covariant formulation applied in a wide variety of contexts. Notable examples include studies of compact stars \cite{Nashed}, anisotropic models \cite{Prad}, neutron stars \cite{Alwan}, and bounce inflation \cite{Hu}. This situation is similar to the case of $f(T)$ models, where a discussion on the original and covariant formulations is also presented \cite{Krssak}, with the spin connection playing a crucial role in the theory.}

{In order to obtain the equations of motion, one performs the variation of the action with respect to the metric, which results in}
\begin{equation}
2f_{QQ} {P^\alpha}_{\mu\nu}\partial_\alpha Q+\frac{1}{2}g_{\mu\nu} (f - Qf_Q )+ f_Q G_{\mu\nu} = 8\pi T_{\mu\nu} \ ,\label{eom1}
\end{equation}
\noindent where $f_Q \equiv df(Q)/dQ$ and $f_{QQ} \equiv d^2f(Q)/dQ^2$.

Since we consider spherical stars, we adopt a spherically symmetric metric of the form
\begin{equation}
    ds^2=e^{A(r)}\, dt^2-e^{B(r)}\, dr^2-r^2\, d\theta^2-r^2 \sin ^2 \theta \, d\phi^2. \label{metric}
\end{equation}

The nonmetricity scalar for a spherically symmetric spacetime then reads

\begin{equation}
  Q(r) = \frac{\left(e^{-B}-1 \right)
  \left( A' + B' \right)} {r},
  \label{Qr}
\end{equation}
where the prime represents the derivative with respect to the radial coordinate $r$.

On the right-hand side of the equations of motion (\ref{eom1}), it is natural to adopt the energy-momentum tensor for a perfect fluid in order to describe compact stars. This assumption leads to the following set of field equations:

\begin{equation}
16\pi r^2\rho e^B = 2rf'_Q(e^B-1)+f_Q[(e^B-1)(2+rA')+(e^B+1)rB']+fr^2e^B
\label{E00}
\end{equation}

\begin{equation}
16\pi r^2 P e^B = -2rf'_Q(e^B-1)-f_Q[(e^B-1)(2+rA'+rB')-2rA']-fr^2e^B
\label{E11}
\end{equation}

\begin{equation}
32\pi r P e^B = 2rf'_QA'-f_Q[2A'(e^B-2)-rA'^2+B'(2e^B+rA')-2rA'']-2fre^B,
\label{E22}
\end{equation}
where $\rho$ and $P$ stand for energy density and pressure, respectively.

By appropriately combining equations (\ref{E00}) and (\ref{E11}), we obtain the following relation:
\begin{eqnarray}
(A' + B')f_Q =8\pi r(\rho+P){\rm e^B}.
\label{AlBlFQ}
\end{eqnarray}

Notice that for vacuum, the above equation reads $A' + B' = 0$, which is identical to the result obtained in General Relativity. With the appropriate redefinition of the time coordinate, one has $A = - B$ for vacuum. As a result, the spacetime outside a spherically symmetric matter distribution in f(Q) gravity coincides with the vacuum solutions of General Relativity. We refer the reader to Refs. \cite{Lin} and \cite{Zhao} for more details. In addition, in Section \ref{Ext} we provide a detailed derivation for the particular $f(Q)$ adopted in this paper.

\section{Modelling stars in $f(Q) = Q +\xi Q^2$ gravity}
\label{sectionMS}

Now, we consider the basic equations to model stars for a particular $f(Q)$ gravity, namely,
\begin{eqnarray}
f(Q)= Q + \xi \, Q^2 \, ,
\label{fQ}
\end{eqnarray}
where $\xi$ is an arbitrary real. This choice represents the simplest extension of $f(Q)$ gravity and is inspired by the Starobinsky model in
$f(R)$ gravity, which adopts the same functional form. However, the approach presented in this paper allows for a straightforward generalization to other functional forms of $f(Q)$. In particular, when $\xi=0$, the results of symmetric teleparallel general relativity are recovered.

From equations (\ref{Qr})-(\ref{E22}), one obtains second-order differential equations for $A$ and $B$, that is, $A''$ and $B''$, which can be numerically solved for a given equation of state (EOS) together with an equation that relates $A'$ and $P'$, derived from the ``conservation equation" of $T_{\mu\nu}$.

Notice that, since the system of equations (\ref{Qr})-(\ref{E22}) does not depend on $A$, one only needs to solve numerically $A''$ to obtain $A'$, that is, a first-order differential equation for $A'$. On the other hand, the numerical integration of $B''$ requires that, in addition to $B'$, $B$ must also be obtained. In Ref. \cite{Lin}, this procedure was followed for a particular EOS.

However, see Ref. \cite{Lin} may be extended and improved in many ways. First, the numerical solution is strongly simplified by means of a new equation which has not been considered so far in the literature. With this new equation, the system to be solved is made up only of first-order differential equations. Second, we study the stellar structure for different polytropic indexes. Third, we discuss and calculate appropriately the mass of the stellar configuration, which also elicits some interesting implications.

\subsection{Interior solution}

As mentioned in the previous section, there is a new equation that substantially simplifies the calculation. It is worth mentioning once again that this equation is new in the literature and, therefore, has not appeared in any article published so far. This new equation is obtained by an algebraic combination of equations (\ref{Qr}) and (\ref{AlBlFQ}), namely,
\begin{eqnarray}
A' + B' = \frac{r \, e^B}{4\xi(e^B-1)}\left[ 1 - \sqrt{1-64\pi\xi(\rho+P)(e^B-1)}\right].
\label{AlBl}
\end{eqnarray}
Thus, instead of integrating $B''$ to obtain $B'(r)$ and $B(r)$, we integrate the above equation for $B'$ to derive $B(r)$. In order to complete the system of equations to be solved numerically, for say a polytropic EOS, one must consider the following equations:
\begin{eqnarray}
    A'' &=& \biggl\{ A'\xi r(e^B-1)(A'+B')[4A'e^B-(A'+B')(3e^B+1)]+ \nonumber \\
    & & -2\xi(e^B-1)^2 (A'+B')[(A'+B')(e^B+5)-2A']+ \nonumber \\
    & & + A'r^2 e^B[(A'+B')(e^B+1)-2A' e^B]+ re^B(e^B-1)(4A' + 6B')+ \nonumber \\
    & & -16\pi r^2 \rho \,e^{2B}[2(e^B-1) + A'r ]\biggr\}\times \nonumber \\
    & & \times
        \biggl\{ 2r(e^B - 1)[2\xi(A'+B')(1-e^B) + r e^B]\biggr\}^{-1} ,
        \label{All}
\end{eqnarray}
which is a first-order differential equation for $A'$, and the ``conservation equation"
\begin{equation}
    2\, P' + (P+\rho)A' = 0
\label{ce1}    
\end{equation}
(see, e.g., \cite{Zhao}).

Before proceeding, it is worth commenting on two consequences of equation (\ref{AlBl}). First, for $\xi \rightarrow 0$, one obtains

\begin{eqnarray}
A' + B' =8\pi(\rho+P){r\, e^B},
\label{AlBlGR}
\end{eqnarray}
which appears, as it should be, in the derivation of the Schwarzschild internal solution.

Second, substituting equation (\ref{AlBl}) into (\ref{Qr}), one obtains the following equation
\begin{eqnarray}
Q(r) = \frac{1}{4\xi}\left[ \sqrt{1-64\pi\xi(\rho+P)(e^B-1)} - 1\right].
\label{QrN}
\end{eqnarray}
Notice that $Q(r) \leq 0$ for any value of $\xi$. Since for regularity at the origin we set $B(0) = 0$, this implies that $Q(0) = 0$. Moreover, outside the matter distribution $Q(r \ge R) = 0$.

Note also that for $\xi \rightarrow 0$, $Q(r)$ reads
\begin{eqnarray}
Q(r) = - 8\pi(\rho+P)(e^B-1).
\label{QrNGR}
\end{eqnarray}

Since $Q(r) = 0$ outside the matter distribution, one could argue that the mass $M$ would be given as in GR, that is, by the integration of
\begin{eqnarray}
\frac{dm}{dr}=4\pi \rho  r^2\, .
\label{dmdr}
\end{eqnarray}

On the other hand, the total rest mass $M_0$ is obtained integrating the following differential equation
\begin{eqnarray}
\frac{dm_0}{dr}=4\pi \rho_0 \,e^{B/2} r^2 \, ,
\label{dm0dr}
\end{eqnarray}
where $\rho_0$ is the rest mass density and $4\pi e^{B/2} r^2 dr$ is the proper volume element. Notice that in bound configurations one has $M < M_0$. This is why it is useful to calculate it.

It should be noted that $M_0$ is unequivocally given by the integration of Equation (\ref{dm0dr}), because it only considers the total rest mass. In this equation, the gravitational and internal energy are not considered.

Concerning the calculation of $M$, the gravitational energy and the internal energy are obviously taken into account. 

Although Equation (\ref{dmdr}) has been adopted in this article and in similar studies, such as those of \cite{Lin} and \cite{Alwan}, it is not guaranteed that $M$ unequivocally represents the mass of a compact object in $f(Q)$ gravity.

In the next subsection, devoted to the exterior solution, we also consider another way to calculate the mass. The detailed issues related to the modeling of stars, in the particular $f(Q)$ adopted in this article, are considered in the next section, where we also compare both star-mass calculations. 

\subsection{Exterior solution}
\label{Ext}

The exterior solution since
\begin{eqnarray}
A' + B' = 0, 
\label{AlBlF}
\end{eqnarray}
is just like the static spherical vacuum solution provided by General Relativity, namely, the Schwarzschild solution. Therefore, the external (vacuum) solution is obviously asymptotically flat. Recall that, without loss of generality, the time coordinate is modified in such a way that
\begin{eqnarray}
A + B = 0.
\label{AB0}
\end{eqnarray}

At this point, it is worth mentioning that we are not imposing or assuming that the exterior solution is given by the Schwarzschild spacetime. Instead, we show in the following that this solution emerges naturally from the equations of motion.

Notice that the vacuum solution is also present in equation (\ref{All}). 
Substituting equation (\ref{AlBlF}) into (\ref{All}) and setting $\rho = 0$, one obtains a differential equation independent of $\xi$, namely
\begin{eqnarray}
A'' = - \frac{A'}{r} - \frac{e^{-A}}{e^{-A}-1} A'^2,
\end{eqnarray}
whose first integral reads
\begin{eqnarray}
e^A\left(1+ rA'\right) = 1 \qquad {\rm or} \qquad (r e^A)' = 1,
\label{Alv}
\end{eqnarray}
where the integration constant is obtained considering $A'=0$ at infinity. Notice that this equation is the same one obtained in the derivation of the Schwarzschild spacetime.

Similarly, from the differential equation for $B''$, which is not shown here, one again finds a differential equation independent of $\xi$, namely,
\begin{eqnarray}
B'' = - \frac{B'}{r} + \frac{e^{B}}{e^{B}-1} B'^2,
\end{eqnarray}
with the first integral, under the condition $B'=0$ at infinity, reading 
\begin{eqnarray}
e^{-B}\left(1- rB'\right) = 1 \qquad {\rm or} \qquad (r e^{-B})' = 1 \ .
\label{Blv}
\end{eqnarray}
Alternatively, the above equation could be obtained using Equation (\ref{AB0}).

The solutions for equations (\ref{Alv}) and (\ref{Blv}) are as follows
\begin{eqnarray}
e^A = e^{-B}  = 1 + \frac{C}{r}\, ,
\end{eqnarray}
where, as in GR, we set $C = -2M_S$. Therefore, we finally have
\begin{eqnarray}
e^A = e^{-B}  = 1 - \frac{2 M_S}{r}.
\label{MS}
\end{eqnarray}

One could ask whether $M_S$ (the mass ``measured by an observer at infinity"), which is ultimately obtained geometrically, also follows from the integration of equation (\ref{dmdr}). This issue related to the calculation of the mass is addressed in Section \ref{Ne}.

From the discussion just above one also sees that it is not necessary to consider the match condition at the surface of the star, since the differential equations for $A''$ and $B''$ contain the vacuum solution as well. It is worth stressing again that this vacuum solution, which is clearly the Schwarzschild spacetime, emerges naturally from the equations of motion studied here; it is not an imposition or a choice.

\section{Numerical examples}
\label{Ne}
In this section, we provide numerical examples of models of polytropic stars in the context of $f(Q)$ gravity, as described by Equation (\ref{fQ}). It is important to note that polytropic EOSs offer a simple and effective way to compare star modeling in GR and $f(Q)$ gravity \cite{eos0}. Furthermore, we investigate the behavior of the nonmetricity scalar $Q(r)$ as well as the metric functions $A(r)$ and $B(r)$, within the same $f(Q)$ framework.

We plan to explore realistic EOS in a future study. As is well known, realistic EOSs can be represented by multiparametric piecewise polytropic equations of state. This is why, as a first approach, the use of polytropic EOSs is of interest.

\subsection{Polytropic stars}

Polytropic equations of state are widely adopted in studies of stellar structure. See \cite{eos0,BS2010} for references to a polytropic approach to neutron stars.

It is well established that polytropic EOS reads
\begin{equation}
    P = k\, {\rho_o}^ \gamma,
    \label{tpol}
\end{equation}
where $P$ is the pressure, $\rho_o$ is the rest-mass density, $k$ is the polytropic gas constant, and $\gamma$ is the polytropic exponent, which is related to the polytropic index $n$ via $\gamma \equiv 1+1/n$. The mass-energy density $\rho$ is easily obtained via the first law of thermodynamics and is given by $\rho = \rho_o + n\,P$.

Notice that in geometrized units, $k^{n/2}$ has unit of length. Consequently, the following dimensionless quantities can be defined: $\bar{r} = k^{-n/2}r $, $\bar{P} = k^{n}P $, $\bar{\rho} = k^{n}\rho $, $\bar{M} = k^{-n/2}M $ and $\bar{Q} = k^{n}Q $ \cite{BS2010}. This is the same as for the set $k = G = c = 1$ in all of our equations. For simplicity, the bars are omitted in our equations.

To model stars in the $f(Q)$ gravity adopted in this article, we basically follow the same procedure as adopted in the modeling of stars in GR. Thus, we need to establish the following central boundary conditions
\begin{equation}
    m = 0  \quad {\rm and}  \quad P = P_c \quad {\rm at} \quad r = 0 \, .
\end{equation}

Furthermore, since it is now necessary to solve the first-order differential equations for $A'$ and $B$, the boundary conditions for these functions must also be established. The regularity conditions in the center imply that $A'(0) =0 $ and $B(0) = 0$. With all these boundary conditions at hand, one integrates the set of differential equations to obtain the structure of the star, i.e. $m(r)$, $P(r)$, and $\rho(r)$. 

The radius $R$ of the star is given by $P(R) = 0$. That is, one starts the integration of the set of differential equations at $r=0$ and continues it to the value of $r$ for which $P(r)=0$. The mass $M  \equiv m(R)$ of the star, since the exterior solution is given by the Schwarzschid metric, is given by the ADM mass, i.e., just like in GR, which is obtained by the integration of equation (\ref{dmdr}).

On the other hand, one could easily obtain the mass via equation (\ref{MS}), since $R$, $A(R)$ and $B(R)$ are known immediately after the numerical integration. In this case, the mass is obtained geometrically. We will see that $M_S$ is not equal to $M$. Thus, one may wonder which of the masses, $M$ or $M_S$, represents the mass of the star. Later, we will discuss this interesting and important issue again.

A usual procedure for comparing the modeling of compact stars in GR with any other alternative theory of gravity is through sequences of ``Mass $\times$ Radius'' and ``Mass $\times$ $\rho_c$'' for a given EOS.

Before presenting the models, it is worth recalling that the speed of sound ($c_{s}$) in units of the speed of light of a polytropic is easily obtained and reads

{
\begin{equation}
    c_{s} = \sqrt{\frac{\gamma -1}{1 + (1 - 1/\gamma)p^{1/\gamma - 1}}}.
    \label{sound}
\end{equation}}
Note that the lower the pressure, the lower the speed of sound. Moreover, with this equation one obtains the causality limit for a given polytropic exponent $\gamma$. In other words, the maximum value of the central density or pressure is obtained.

From equation (\ref{sound}), one concludes that if $\gamma \leq 2$, the central density or pressure can be arbitrarily large. This means that polytropics with $\gamma \leq 2$ never exceed the causality limit. On the other hand, for $\gamma > 2$, there is a maximum central density allowed, otherwise the causality limit is exceeded.

{Another important issue to consider before proceeding is related to the energy conditions. Recall that they are the null (NEC), weak (WEC), dominant (DEC), and strong (SEC) energy conditions, which are given by:}
{
\begin{itemize}
    \item NEC: $\rho + p \ge 0$
    \item SEC: $\rho + p \ge 0$, $\rho + 3p \ge 0$
    \item DEC: $\rho \ge 0$, $\rho - p \ge 0$
    \item WEC: $\rho \ge 0$, $\rho + p \ge 0$
\end{itemize}
}
{For polytropic EOSs, NEC, SEC and WEC are obviously satisfied. However, this is not the case for DEC. Notice that the combination of equation (\ref{tpol}) with $\rho = \rho_o + n\,P$ provides}
{
\begin{equation}
    \rho - p = (n-1)p+p^{n/(n+1)}.
\label{DEC}
\end{equation}
}
{From this equation it is readily concluded that for $n \ge 1$ (or $\gamma \le 2$) DEC is satisfied. However, for $n < 1$ (or $\gamma > 2$), depending on the values of $\rho$, it is possible to have $\rho - p < 0$, which implies that DEC is violated.}

In the first set of models, we consider a polytropic EOS with index $n = 1$, which gives $P= \rho_0^2$, where we set $k=1$ since, as already mentioned, we are dealing with dimensionless quantities.

\begin{figure}[h]
    \centering
    \begin{subfigure}[b]{0.49\textwidth}
        \centering
        \includegraphics[width=\textwidth]{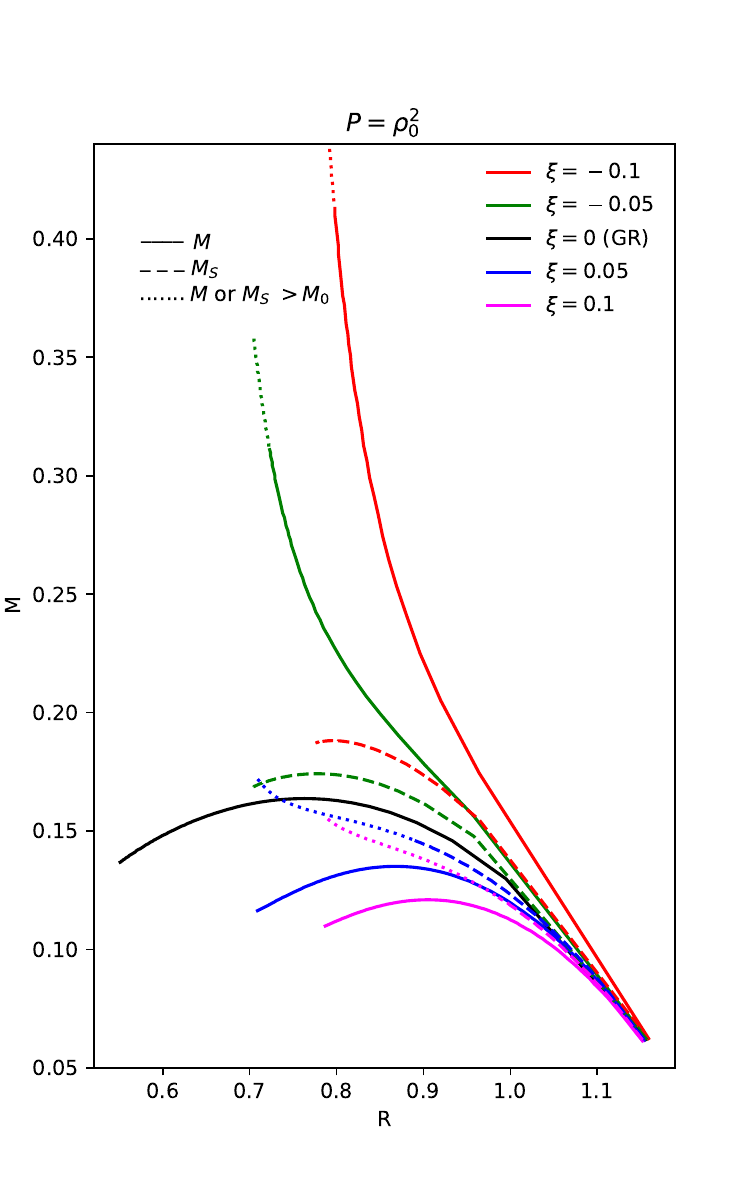}
        \label{N1e}
    \end{subfigure}
    \hfill 
    \begin{subfigure}[b]{0.49\textwidth}
        \centering
        \includegraphics[width=\textwidth]{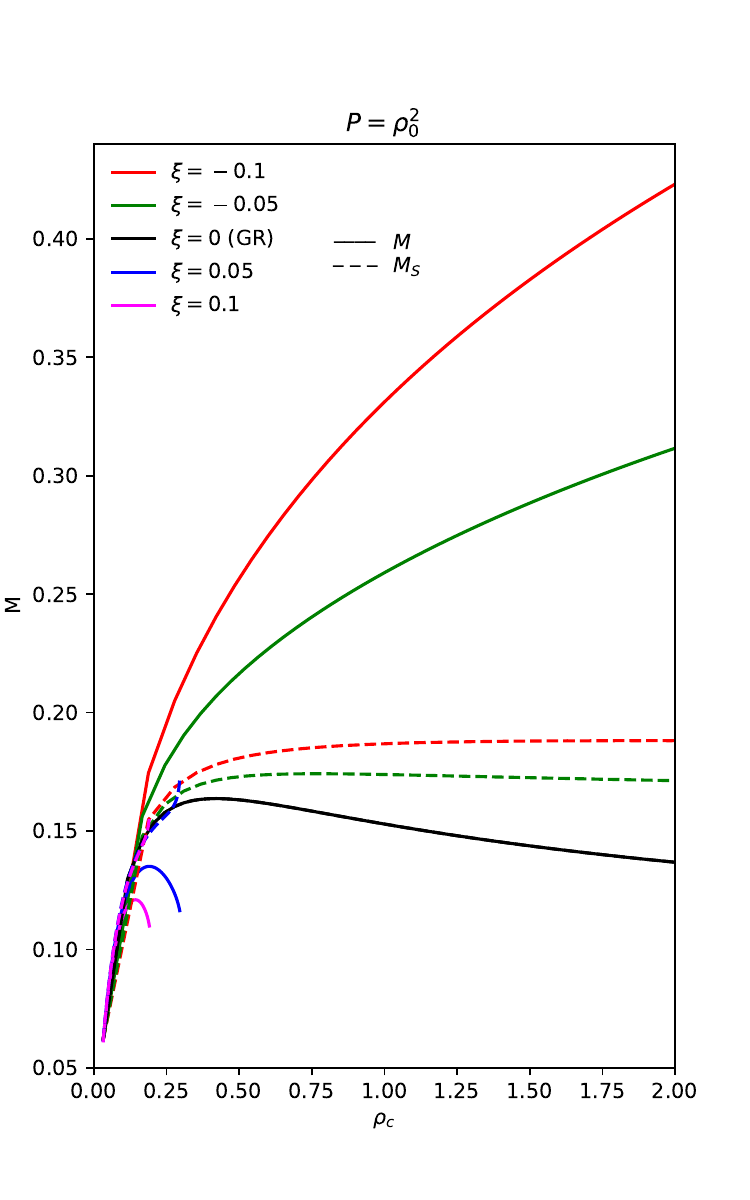}
        \label{N1d}
    \end{subfigure}
    \caption{Left (Right): sequences of M and M$_S$ vs. radius $R$ (central mass-energy density $\rho_c$) for $P= \rho_0^2$ and different values of $\xi${, namely, $- 0.1$ (red lines), $- 0.05$ (green lines), 0 (black line), $0.05$ (blue lines), and $0.1$ (magenta lines).} {All quantities are shown in dimensionless units (with $G = c = k = 1$)}.}
    \label{N1}
\end{figure}

In Figure \ref{N1}, we see that for a range of values of $\rho_c$, one obtains the corresponding curves ``Mass $\times$ Radius'' and ``Mass $\times$ $\rho_c$''. 
From these curves, we can identify, for example, the maximum mass allowed for a given EOS. For $\xi=0$, we have $f(Q)=Q$, which is nothing more than STGR, which is equivalent to GR. One can also see from the figure how the way of calculating the total mass affects the results. The continuous lines represent the curves obtained using the ADM mass, and the dashed lines represent the curves obtained using $M_{S}$.

From the curves in Figure \ref{N1}, where $n=1$ ($\gamma =2$), it is clear that, for $\xi < 0$, there are maximum masses for $M_S$, which are greater than that for GR. Note also that the curves for $M_S$ have shapes similar to those for GR.
The same does not occur for $M$, i.e., there is apparently no maximum mass for $M$. In addition, the masses for a given $\rho_c$ increase with decreasing values of $\xi$. For low densities, for say $\rho_c < 0.1$, $M_S \sim M$. However, for $\rho_c > 0.1$,  $M > M_S$, and the higher the density, the greater the difference between $M_S$ and $M$.

For $\xi > 0$, there are maximum masses for $M$, which are lower than those for GR, and the higher $\xi$, the lower the maximum masses. For $M_S$, in this case, it was not possible to identify maximum masses. Furthermore, it was not possible to follow the calculations for densities above the values shown in Figure \ref{N1} due to numerical instabilities.

Furthermore, there is a condition to be obeyed, that is, $M$ and $M_S$ must always be smaller than the total rest mass $M_0$, given in Equation (\ref{dm0dr}), for the configuration to be bound. Therefore, we identified the intervals where this is violated by the dotted lines in the left panel of Figure \ref{N1}. Taking this condition into account, one concludes that both $M$ and $M_S$ indeed have maxima; otherwise, the configurations would not be bound.

Although for GR no dotted line is shown, this violation also occurs for values of $R$ smaller than those shown in the left panel of Figure \ref{N1}. This is not a problem for GR, as the region where the violation occurs is well known to be dynamically unstable to radial perturbations \cite{Shapiro}.

\begin{figure}[h]
    \centering
    \begin{subfigure}[b]{0.49\textwidth}
        \centering
        \includegraphics[width=\textwidth]{{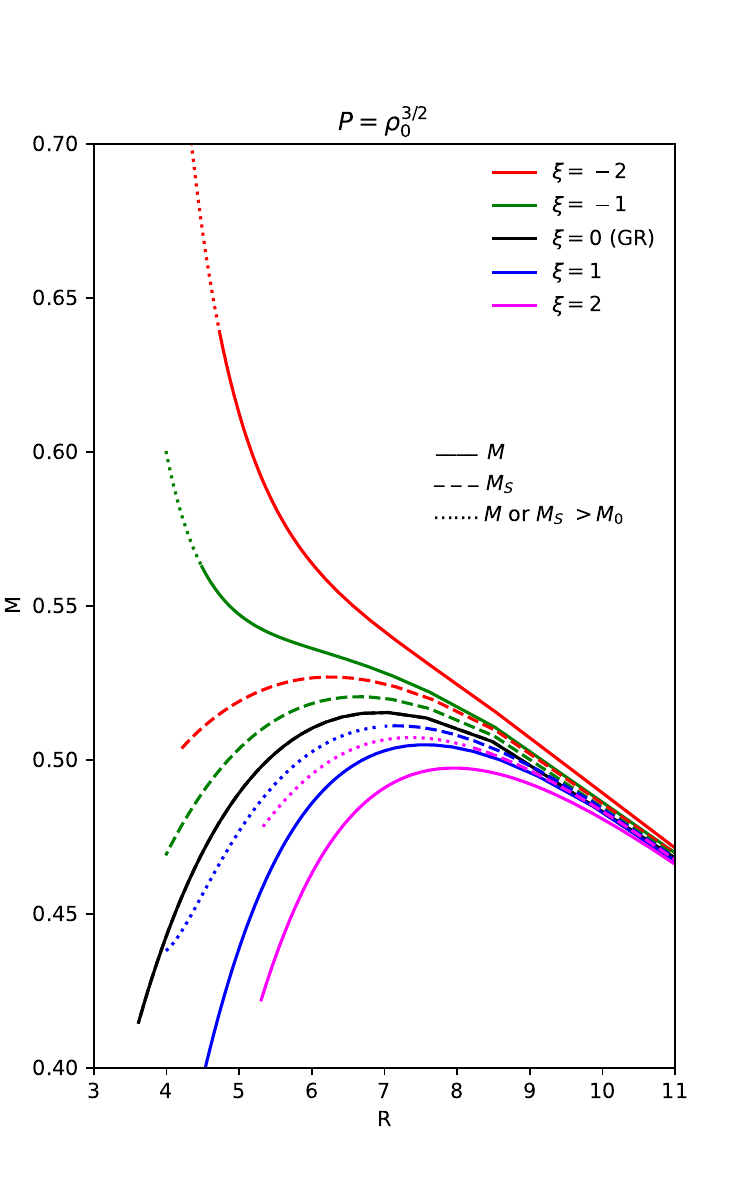}}
        \label{N2e}
    \end{subfigure}
    \hfill 
    \begin{subfigure}[b]{0.49\textwidth}
        \centering
        \includegraphics[width=\textwidth]{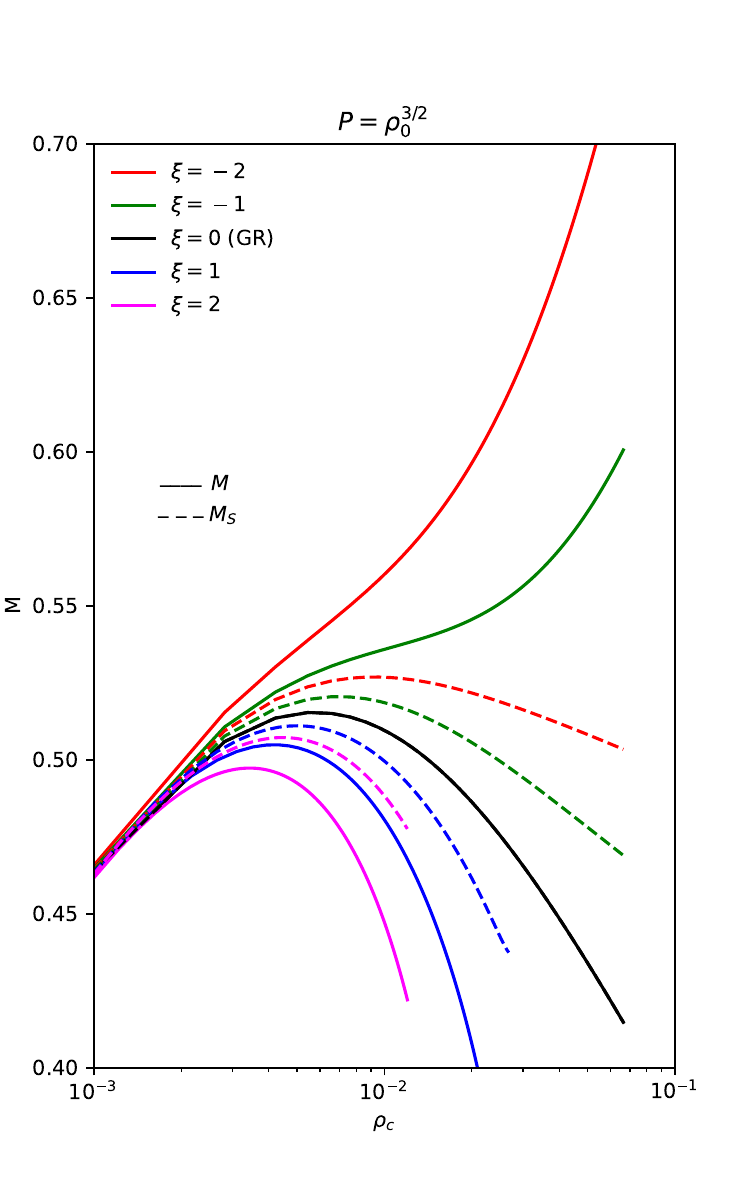}
        \label{N2d}
    \end{subfigure}
    \caption{{Left (Right): sequences of M and M$_S$ vs. radius $R$ (central mass-energy density $\rho_c$) for $P= \rho_0^{3/2}$ and different values of $\xi${, namely, $- 2$ (red lines), $- 2$ (green lines), 0 (black line), $1$ (blue lines), and $2$ (magenta lines).} {All quantities are shown in dimensionless units (with $G = c = k = 1$)}.}}
    \label{N2}
\end{figure}

In Figure \ref{N2}, we have considered the polytropic EOS with $n=2$ {($\gamma = 3/2$), which is softer than $n =1$.} In this case, the difference between the curves for different $\xi$'s becomes clearer. For $\xi \ge 0$, there are maximum masses for both $M$ and $M_S$, which are lower than for GR. In addition, most of the curve lies within the violation region of the condition $M_0>M_S$. For $\xi<0$, it is clear that we have the maximum for $M_S$ and that they are greater than the maximum for GR. The maximum for $M$ comes from the condition $M = M_0$, as well as from the previous case $n=1$.

We also consider models for a polytropic EOS stiffer than $n = 1$ and $n=2$, in particular for $n = 2/3$ ($\gamma = 5/2$). In this case, there is a maximum value for the central density due to the causality limit. Using Equation (\ref{sound}), one obtains such a limit, namely

\begin{equation}
    \rho_{max} = \frac{3}{2} \left( \frac{6}{5}\right)^{5/3} \simeq 2.03 \ ,
\end{equation}

\par\noindent
which is well above the values adopted in our calculation. {Concerning DEC, it is readily obtained from equation (\ref{DEC}), that it is violated for} 

\begin{figure}[h]
    \centering
    \begin{subfigure}[b]{0.49\textwidth}
        \centering
        \includegraphics[width=\textwidth]{{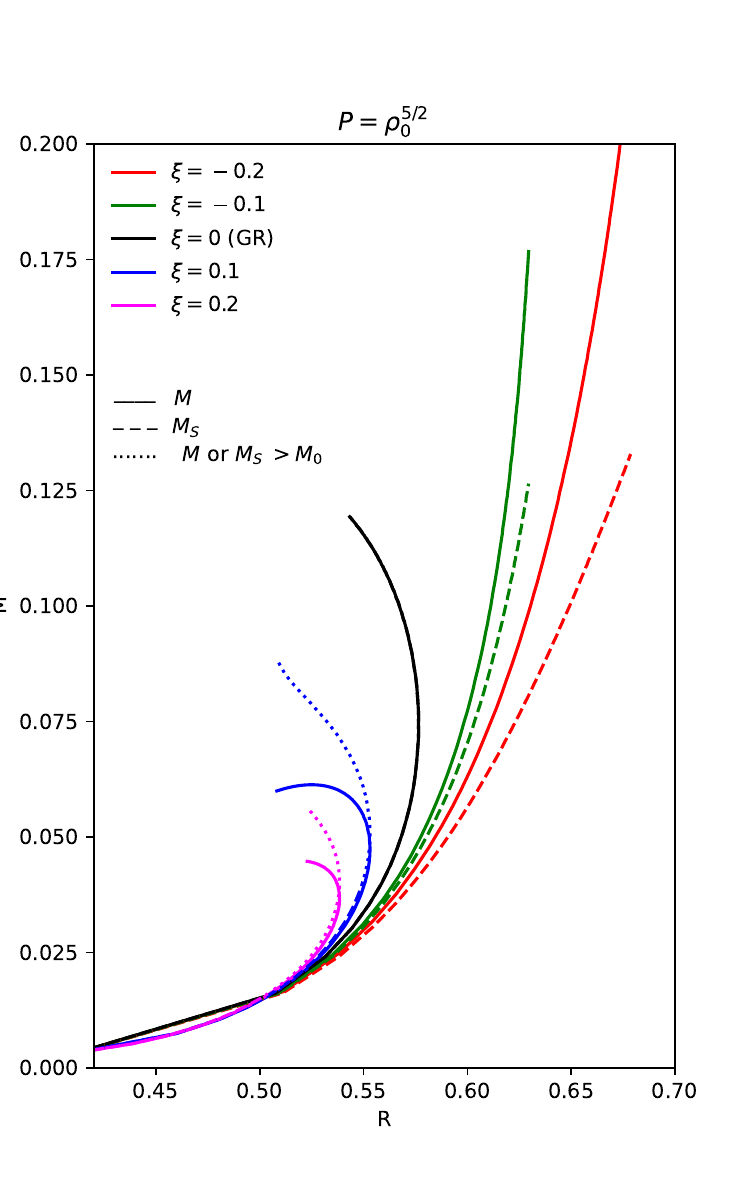}}
        \label{N23e}
    \end{subfigure}
    \hfill 
    \begin{subfigure}[b]{0.49\textwidth}
        \centering
        \includegraphics[width=\textwidth]{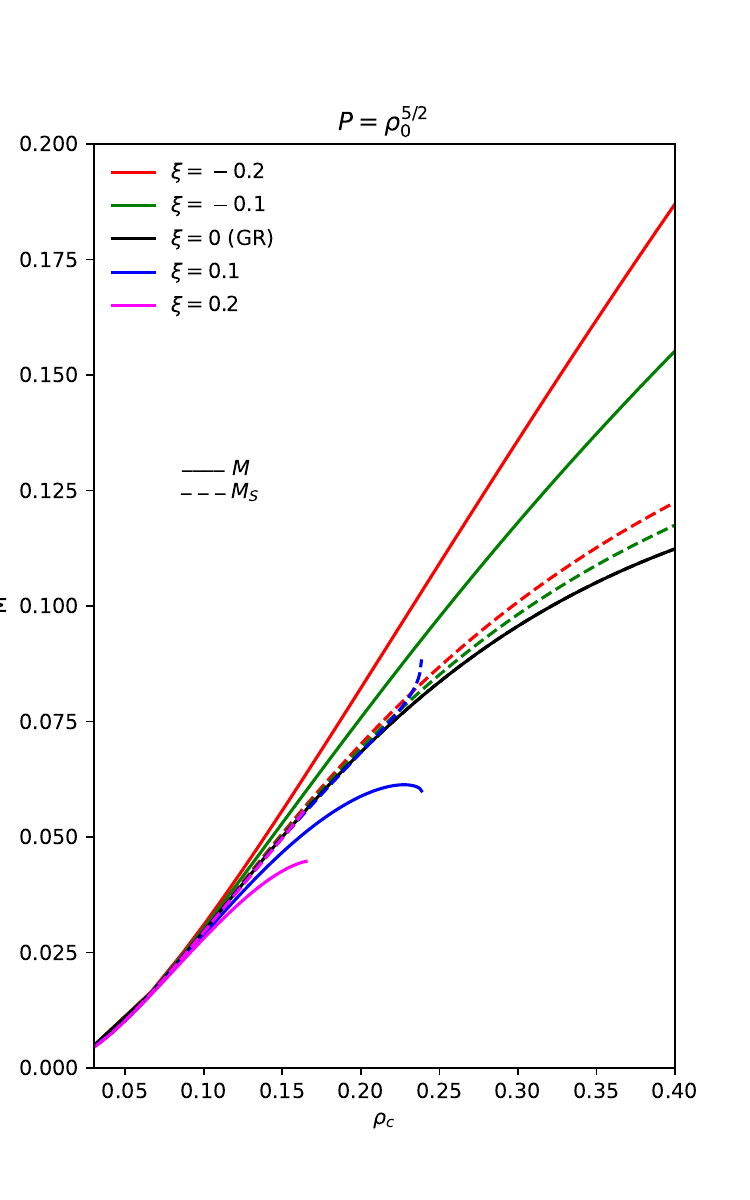}
        \label{N23d}
    \end{subfigure}
    \caption{{Left (Right): sequences of M and M$_S$ vs. radius $R$ (central mass-energy density $\rho_c$) for $P= \rho_0^{5/2}$ and different values of $\xi${, namely, $- 0.2$ (red lines), $- 0.1$ (green lines), 0 (black line), $0.1$ (blue lines), and $0.2$ (magenta lines).} All quantities are shown in dimensionless units (with $G = c = k = 1$).}}
    \label{N23}
\end{figure}

{
\begin{equation}
        \rho > 3^{5/3} \simeq 6.24.
\end{equation}
}
Therefore, all models presented in Figure \ref{N23} respect the causality limit and satisfy DEC and all other energy conditions.

As shown in Figure \ref{N23}, qualitatively, the sequences have a behavior similar to the GR models. For $\xi \leq 0$, it is known that the mass, whether $M$ or $M_S$, behaves somewhat differently compared to the cases discussed earlier. Note that for a larger radius, we generally have greater masses. On the other hand, in the cases $n=1$ and $n=2$, the mass decreases significantly with a larger radius. Again, we have most of the curve for $\xi>0$ within the violation region of condition $M_0>M_S$. Meanwhile, for negative $\xi$ values, this violation does not even appear for the density range considered in Figure \ref{N23}.

At this point, it is interesting to mention some general characteristics of the various models presented. Our calculations for different values of $\gamma$ have some features in common. First, for $\xi \leq 0$, stable numerical solutions are obtained for any value of $\rho_c$. Second, for $\xi > 0$ numerical instabilities arise for densities higher than those shown in figures \ref{N1}, \ref{N2} and \ref{N23}.

Before concluding this section, it is worth mentioning that although polytropic EOSs represent a very simplified model for compact star EOSs, they are very useful for different reasons. First, it is an easy way to compare alternative theories with GR. Second, they can be useful for describing realistic EOSs, which can be written in a piecewise polytropic form.

Last but not least, an important issue has to do with the stability of stars in $f(Q)$ gravity. This is a very relevant question that deserves to be appropriately addressed. This issue alone deserves a specific paper because of its complexity. Therefore, we will consider this issue in a future publication.

Left (Right): sequences of M and M$_S$ vs radius $R$ (central mass-energy density $\rho_c$) for $P= \rho_0^{5/2}$ and different values of $\xi$, namely $- 2$ (red lines), $- 2$ (green lines), 0 (black line), $1$ (blue lines) and $2$ (magenta lines). All quantities are shown in dimensionless units (with $G = c = k = 1$).

\subsection{Nonmetricity}

{It is also worthwhile to consider, for a given EOS and different values of $\xi$, the behaviors of the nonmetricity scalar $Q(r)$ and the metric functions $A(r)$ and $B(r)$.}

As already mentioned in Section \ref{sectionMS}, the nonmetricity scalar $Q(r)$ is negative inside the star and null outside the matter distribution, as can be seen by Equation (\ref{QrN}).

From equation (\ref{fQ}), one sees that the more negative $\xi$ is, the more negative $f(Q)$ is and, therefore, the more intense the gravitational interaction. On the other hand, for positive values of $\xi$, the gravitational interaction is less intense.

\begin{figure}
\centering
\includegraphics[scale=0.49]{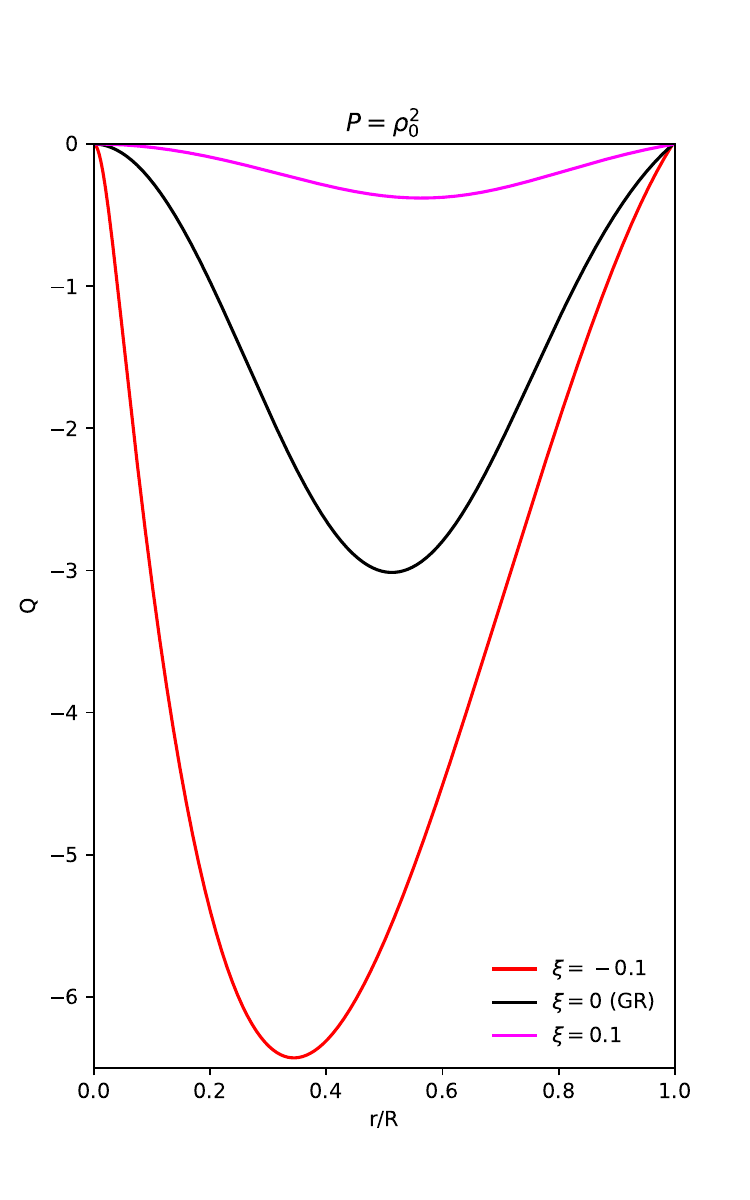}
\caption{Q(r) for maximum $M_S$ of $\xi = - 0.1, 0$ and $0.1$ for $P= \rho_0^2$.}
    \label{Q_r}
\end{figure}

In Figure \ref{Q_r}, $Q(r)$ is shown for the maximum $M_S$ of $\xi = - 0.1$, 0 and 0.1 for the polytropic EOS $P= \rho_0^2$. Notice that the nonmetricity scalar $Q(r)$ has a shape that resembles a ``potential well".

We have considered the behavior of the nonmetricity for the maximum masses, because it is more pronounced. For masses smaller than the maximum masses for a given $\xi$, the nonmetricity curves would be above the magenta ($\xi = 0.1$), black ($\xi = 0$) and red ($\xi = - 0.1$) curves (see Figure \ref{Q_r}), as expected. The ``potential wells" are shallower for masses below the maximum masses.

Regarding metric functions, it is worth noting that the system of equations solved to model the $f(Q)$ stars does not depend on $A(r)$. Therefore, it is not necessary to know this function to model the stars. In any case, we study the behavior of $A(r)$, in addition to B(r).

Before proceeding, it is worth mentioning that the equations used in this paper to model stars are also valid outside the matter distribution since $P$ and $\rho$ go smoothly to zero. Consequently, there are no jumps in the potentials $A$ and $B$ at least in their first derivatives at $r = R$.

However, to obtain $A(r)$, one sees that it is defined up to a constant, as in GR. This is so because $A(r=0)$ is not known before hand. This means that $A(0)$ is obtained after the numerical integration of the system of equations. Since the vacuum is given by the Schwarzschild solution, the constant is obtained by imposing that $A(r)$ matches smoothly onto the Schwarzschild metric at the surface, namely,
\begin{equation}
A(R) = \ln \left( 1 - \frac{2M_S}{R}\right).
\end{equation}

Concerning $B(r)$, it is completely given, since it depends only on its value at $r = 0$ which, by regularity condition, implies that $B(0) =0$, as already mentioned. This is so thanks to equation (\ref{AlBl}) which provides a first-order differential equation instead of a second-order one.

\begin{figure}[h]
    \centering
    \begin{subfigure}[b]{0.49\textwidth}
        \centering
        \includegraphics[width=\textwidth]{{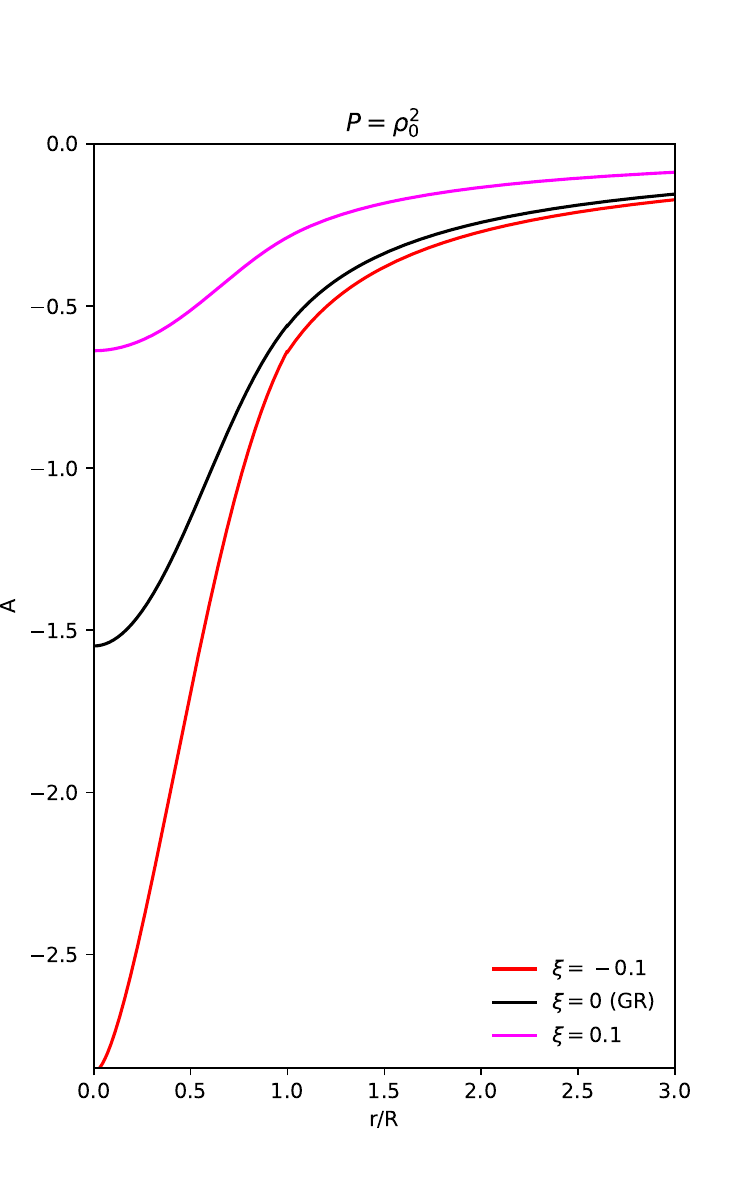}}
        \label{a_r}
    \end{subfigure}
    \hfill 
    \begin{subfigure}[b]{0.49\textwidth}
        \centering
        \includegraphics[width=\textwidth]{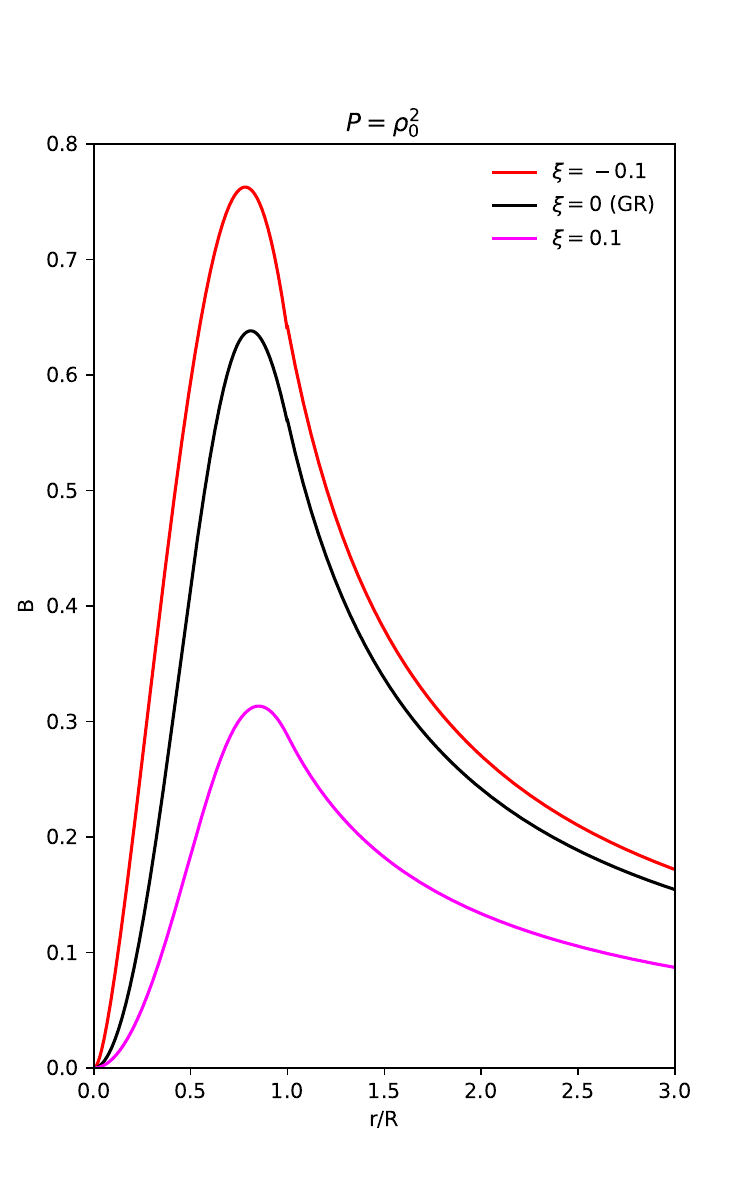}
        \label{b_r}
    \end{subfigure}
    \caption{$A(r)$ and $B(r)$ for maximum $M_S$ of $\xi = - 0.1, 0$ and $0.1$ for $P= \rho_0^2$. {The plots are shown as a function of the normalized radial coordinate $r/R$, with $R$ being the stellar radius. The range is extended beyond the surface ($r/R > 1$) to show the smooth matching between the interior and exterior solutions.}
}
    \label{ab_r}
\end{figure}

We show in Figure \ref{ab_r} the metric functions $A(r)$ and $B(r)$ inside and outside the matter distribution for a polytropic EOS $P= \rho_0^2$ for the maximum $M_S$ of $\xi = - 0.1$, 0 and 0.1. Note that $A(r)$ for $\xi = 0.1$ ($- 0.1$) is above (below) the GR curve. Regarding $B(r)$ for $\xi = - 0.1$ ($0.1$), it is above (below) the GR curve.

{\subsection{A simple polytropic model for neutron stars}}
The aim of this paper and similar ones applied to other modified gravity {is to develop a framework work} to study the structure of neutron stars. This is provided by means of obtaining the TOV equations in these modified gravity theories.

It is well known that ordinary stars and even white dwarfs can be modeled using Newtonian gravity, since they are not compact enough to require the use of general relativity or any modified gravity.

To model neutron stars, it is necessary to know the EOS, which describes the microphysics at extremely high densities occurring deep inside these stars. It is worth mentioning that the detailed knowledge of the neutron star EOS is still an open problem, that is why the literature has a lot of proposals for the so-called realistic EOSs. 

{It is well known that a simple polytropic EOS cannot provide models in agreement with the data coming from observations. This type of modeling can only provide, at most, the correct orders of magnitude for masses, radii, and densities.}

{In a forthcoming paper, to appear elsewhere, we consider some realistic EOSs in the modeling of neutron stars in the context of $f(Q)$ gravity. In such a paper, we apply the same basic set of equations derived here, which provides a robust framework to model neutron stars in $f(Q)$.}

Notice that in previous subsections models are presented without modeling realistic stars, since the main idea was to study the effects of the quadratic additional term ($Q^2$) compared to the models that adopt general relativity. For this purpose, the adoption of polytropic EOSs is extremely useful, as we have seen.

{Having in mind the limitations of polytropic EOSs with respect to agreement with observational data, we restore the constants $k$, $G$ and $c$ in our equations, in order to obtain a toy model in the particular $f(Q)$ studied in  this paper.}

 We present, in particular, models for a polytropic EOS with index $n = 1$. To proceed, it is necessary to set the value of the polytropic gas constant, $k$. {This constant is not unique for a given value of polytropic index n.}
For $n =1$ this constant can be given by $k \simeq 2.0\times10^5$ cm$^5$/s$^2$/g (see, e.g.,\cite{Prakash}), for an interacting Fermi gas.

\begin{figure}[h]
    \centering
    \begin{subfigure}[b]{0.49\textwidth}
        \centering
        \includegraphics[width=\textwidth]{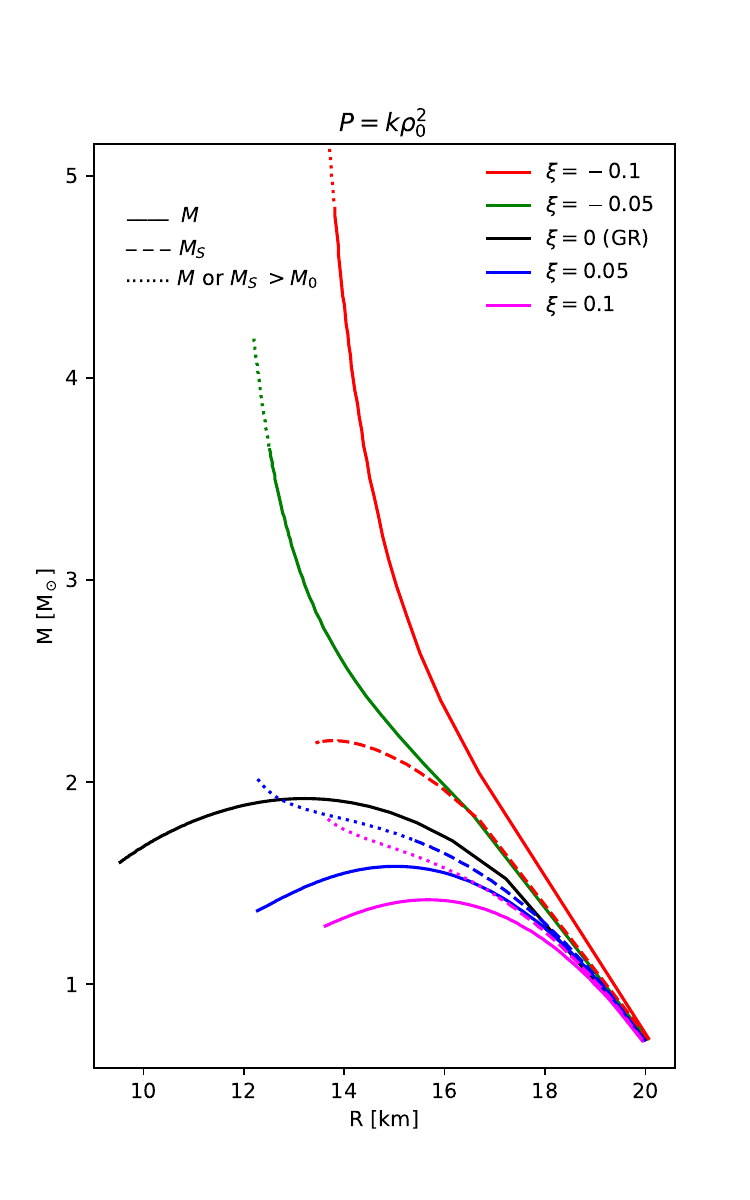}
        \label{N1eunits}
    \end{subfigure}
    \hfill 
    \begin{subfigure}[b]{0.49\textwidth}
        \centering
        \includegraphics[width=\textwidth]{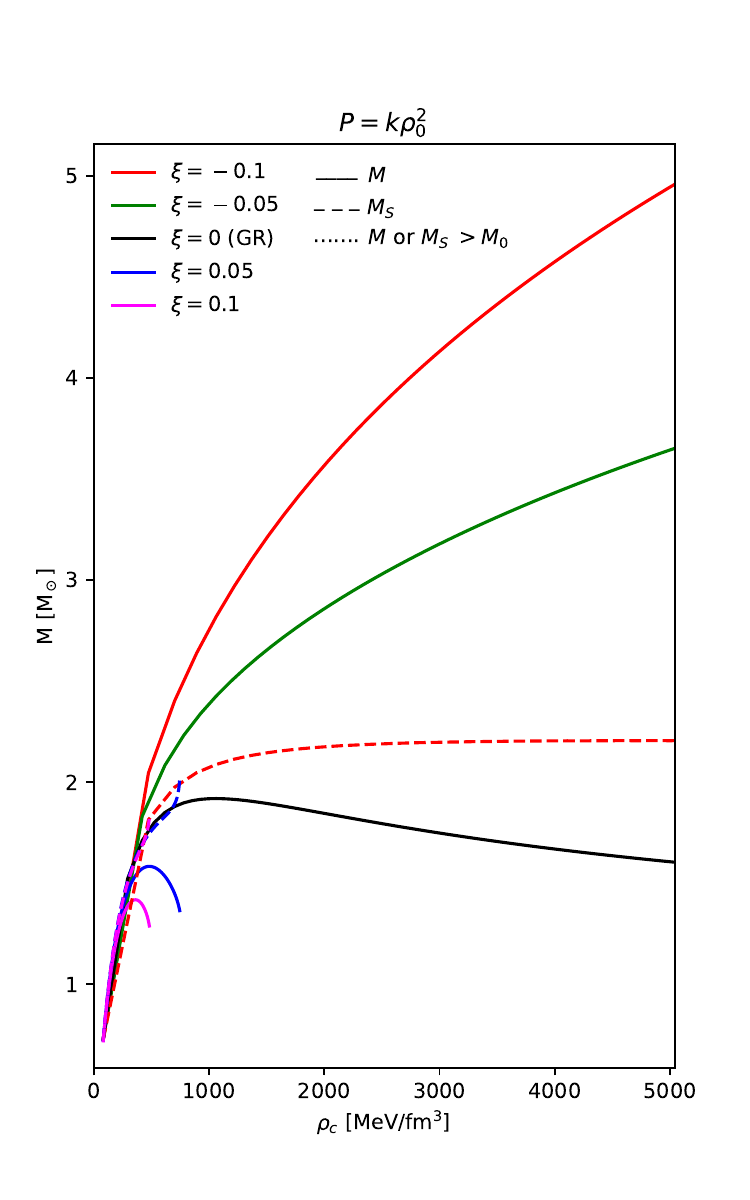 }
        \label{N1dunits}
    \end{subfigure}
    \caption{{Left (Right): sequences of M and M$_S$ vs. radius $R$ (central mass-energy density $\rho_c$) for $P= k\rho_0^2$ and different values of $\xi${, namely, $- 0.1$ (red lines), $- 0.05$ (green lines), 0 (black line), $0.05$ (blue lines), and $0.1$ (magenta lines).}}}
    \label{N1units}
\end{figure}

{In Figure \ref{N1units}, we present sequences of ``Mass $\times$ Radius'' and ``Mass $\times$ $\rho_c$'' for $n=1$ now with the usual units. Note that although it is a very simple model for an EOS, it provides models with figures similar to some realistic EOSs, in particular with respect to mass and energy densities. Concerning the radii, they are greater than usually predicted even for the stiffest realistic EOSs.}

{Even with their inherent limitations, some interesting conclusions can be obtained from polytropic models. Notice that, for example, depending on the value of $\xi$, maximum masses much greater than 2 M$_\odot$ can be obtained.}

{These results indicate that when realistic EOSs are adopted, the maximum mass in $f(Q)$ gravity can be much greater than those predicted by general relativity models. As already mentioned, in a forthcoming paper we consider such issues in detail.}

{Before concluding this subsection, it is worth commenting on the consequences of considering different values of $k$. Note that $M$ and $R$ are directly proportional to $\sqrt{k}$, whereas $\rho \propto 1/k$. Therefore, a value of $k$ greater (lower) than that of Figure \ref{N1units} would shift, for example, the ``Mass $\times$ Radius'' curves upward (downward) and to the right (left). Consequently, the larger (smaller) $k$ is, the larger (smaller) the masses and radii will be.}

\vspace{0.5cm}
\section{Final remarks}
\label{Fr}

{In this paper, we have presented and discussed the implications of considering $f(Q)=Q+\xi \, Q^2$ for both different values of $\xi$ and different polytropic EOSs.}

{The polytropic EOSs allow us to effectively compare alternative models with GR and are a previous step before considering realistic equations of state, which in turn can be written in piecewise polytropic form.}

In general, it can be observed that considering the EOSs to be softer (Figure \ref{N2}) or stiffer (Figure \ref{N23}) significantly alters the behavior of the curves (models).

With respect to the parameter $\xi$, different values were taken. It was clear that this choice has a great influence on the model's behavior, especially with regard to the sign of $\xi$. In general, negative $\xi$'s lead to stable numerical solutions that allow greater masses, either $M$ or $M_S$. For positive values, we have more restrictive models with lower masses and numerical instabilities starting at a given density.

In GR it is well known what the stability limits are for modeling compact objects \cite{Shapiro}. The maximum mass itself defines such a limit. However, we do not know if the same occurs for $f(Q)$. Is there instability even before the mass assumes the value of $M_0$? This question is of great relevance and requires a more in-depth study of the issues involving the absolute stability of a gravity model for compact objects. This topic is under investigation for future work.

{Although a detailed perturbative stability analysis in $f(Q)$ gravity is beyond the scope of this work, it is useful to recall that, in General Relativity, the onset of radial instability in compact stars typically occurs at the turning point where $dM/d\rho_c = 0$ \cite{Shapiro}. This condition indicates the transition from stable to unstable stellar configurations under radial perturbations. Although this criterion has not yet been rigorously established in the context of $f(Q)$ gravity, similar behaviors are expected and have been qualitatively observed in several modified gravity theories, for example, in $f(Q)$ models for strange stars \cite{Lohakare} and Hybrid Stars \cite{Bhar}, and also for compact stars in $f(Q,T)$ gravity \cite{Nashed}. Therefore, in analogy with GR, we associate the maximum mass points in our models with stability thresholds. A more rigorous dynamical analysis will be addressed in future work.}

In this paper, we have considered two ways of calculating the mass of the compact star, namely the ADM mass $M$ given in (\ref{dmdr}) also used in GR and the Schwarzschild mass $M_S$ defined in (\ref{MS}). Although both masses are equal in GR, this is not necessarily the case in alternative gravity theories. In $f(Q)$, we can see that they diverge especially for larger masses. In this sense, it is necessary to evaluate which mass better describes the stellar configuration. We can say that the Schwarzschild mass is more suitable because it is precisely the mass that an observer would see from outside the star. $M_S$ describes the spacetime witnessed by the observer due to the curvature caused by the mass of the star.

In short, $M_S$ seems to consider the effect of the quadratic term in $Q$ more appropriately than the ADM mass. The last one seems not to adequately take into account the additional effects caused by the quadratic term in $Q$. The difference $M_{S} - M_0$, a ``mass defect", gives the bind energy, i.e., the gravitational potential energy in any theory of gravity. 

However, this issue related to the calculation of the mass in different theories of gravity deserves close scrutiny, since in many of them the ADM mass seems not to be the most appropriate way to calculate the mass. It is worth stressing again
that the geometric mass seems to be the more appropriately one. Regarding $f(Q)$, this paper is the first in the literature to address such an issue, which seems to be an open question.

It is worth highlighting the importance of obtaining and using equation (\ref{AlBl}) to numerically solve the problem. This equation was obtained by manipulating (\ref{Qr}) and (\ref{AlBlFQ}) and then solving it as a polynomial equation in $A'+B'$. Without it, the calculations would become more extensive and perhaps not even feasible with the same generality. In the literature, some authors adopt particular functional forms for $A$ and/or $B$ to make the system of differential equations easier to solve. On the other hand, in this paper, Equation (\ref{AlBl}) allows us to simplify the system and solve it numerically without loss of generality.

Finally, it is worth highlighting some novel contributions that this article brings to the literature. In particular, we focus on two closely related studies that adopt the covariant formulation and address the modeling of stars within the same $f(Q)$ gravity framework considered here. 

In Ref. \cite{Lin}, the authors present models for a specific pseudo-polytropic equation of state. In this work, we successfully develop models for polytropic EOSs with varying polytropic indices. This approach is particularly interesting, as it allows us to verify whether stable numerical solutions to the TOV equations can be obtained for the specific $f(Q)$ gravity considered here, for different EOSs.

{Among the various references we cite in this paper within the scope of the $f(Q)$ approach, we highlight two works \cite{Alwan,Lin} from recent literature that deal with the same quadratic functional form $f(Q)=Q+\xi Q^2$. More specifically, in Ref. \cite{Alwan}, some results for realistic equations of state (EOSs) are presented. However, due to the numerical approach employed, they are unable to model cases with $\xi <0$. In contrast, our approach allows for modeling with positive and negative values of $\xi$. This is made possible by our new equation, which significantly simplifies the numerical calculations, providing a robust framework that allows the study of models with any value of $\xi$. In \cite{Lin} the authors also numerically solve the equations for compact stars with the same $f(Q)$, but they consider only the polytropic exponent $\gamma=2$. Furthermore, the numerical solution could be strongly simplified using the new equation (\ref{AlBl}).}

{Although the present work focuses on simple polytropic EOSs, we plan to apply our method to realistic piecewise-polytropic EOSs in future investigations. These EOSs, typically constructed from nuclear physics data, are inherently more complex and pose numerical challenges when solving the TOV equations, especially in theories beyond General Relativity. However, the new equation (\ref{AlBl}) introduced in this work can significantly simplify integration even for these complex EOSs, providing a stable and efficient framework for modeling compact stars with realistic microphysics within the context of $f(Q)$ gravity.}

{It is worth also mentioning that both Refs. \cite{Lin} and \cite{Alwan} calculate the mass of the star in the same way as in TOV GR. We do the same here, but we raise the point that perhaps this is an open question, since it is necessary to prove that such a calculation appropriately takes into account the $f(Q)$ gravity effects. In addition, we propose a way of calculating the mass, which is related to aspects linked to the geometry of space-time. Furthermore, these authors, unlike us, do not check whether the mass of the star is less than $M_0$, which is an essential condition that must be fulfilled}.

\section*{Data availability statement}
In this study, no new data was created or analyzed.
 
\section*{Acknowledgment}
J.C.N.A. thanks CNPq (307803/2022-8) for partial financial support. H.G.M.F. thanks CNPq for financial support (152326/2022-7). Last but not least, we thank the referees for the detailed and valuable reviews of our article, which helped substantially improve it.

\appendix 

\section{Derivation of Equation (\ref{AlBl})}

{First, we need to add equations (\ref{E00}) and (\ref{E11}). After the appropriate cancelations, one obtains the following:
\begin{eqnarray}
    8\pi r(\rho +P)e^B=(A'+B')\, f_Q.
    \label{app1}
\end{eqnarray}}

{Since $f(Q)=Q+\xi Q^2$, we have $f_Q=1+2\, \xi Q$, where
\begin{equation}
  Q(r) = \frac{\left(e^{-B}-1 \right)
  \left( A' + B' \right)} {r}\ .
\end{equation}}

{By substituting this into equation (\ref{app1}) and rearranging the terms, the equality can be expressed as follows:
\begin{eqnarray}
    2\xi (e^{B}-1)(A'+B')^2-re^{B}(A'+B')+(\rho +P)8\pi r^2 e^{2B}=0
\end{eqnarray}}

{We can now solve the previous equation as a quadratic in $(A'+B')$. The solution is:
\begin{eqnarray}
A' + B' = \frac{r \, e^B}{4\xi(e^B-1)}\left[ 1 \pm \sqrt{1-64\pi\xi(\rho+P)(e^B-1)}\right].
\end{eqnarray}}
{The negative sign within the square root must be chosen, since for $\rho+P\rightarrow 0$ we have $A'+B'=0$ according to equation (\ref{AlBlFQ}). Thus, we finally obtain equation (\ref{AlBl}).}

\end{document}